\documentclass[aps,letterpaper,11pt,twoside,tightenlines,nofootinbib,showpacs,preprint,onecolumn]{revtex4}
\usepackage{graphicx}
\usepackage[sort&compress]{natbib}
\usepackage{latexsym}
\usepackage{epsfig}
\begin{document}
\newcommand{\eg}{{\it e.g.}}
\newcommand{\etal}{{\it et. al.}}
\newcommand{\ie}{{\it i.e.}}
\newcommand{\be}{\begin{equation}}
\newcommand{\dd}{\displaystyle}
\newcommand{\ee}{\end{equation}}
\newcommand{\bea}{\begin{eqnarray}}
\newcommand{\eea}{\end{eqnarray}}
\newcommand{\bef}{\begin{figure}}
\newcommand{\eef}{\end{figure}}
\newcommand{\bce}{\begin{center}}
\newcommand{\ece}{\end{center}}
\def\lsim{\mathrel{\rlap{\lower4pt\hbox{\hskip1pt$\sim$}}
    \raise1pt\hbox{$<$}}}         
\def\gsim{\mathrel{\rlap{\lower4pt\hbox{\hskip1pt$\sim$}}
    \raise1pt\hbox{$>$}}}         

\title{Spontaneous breaking of translational invariance in 
non-commutative $\lambda \phi^4$ theory in two dimensions. }

\author{P. Castorina}
\altaffiliation{Email address: paolo.castorina@ct.infn.it}

\affiliation{Department of Physics, University of Catania and INFN Sezione di 
Catania, Via S. Sofia 64, I 95123,  Catania, Italy}

\author{D. Zappal\`a}
\altaffiliation{Email address: dario.zappala@ct.infn.it}

\affiliation{INFN Sezione di Catania, Via S. Sofia 64, I 95123,  Catania, Italy}

\date{\today}
\begin{abstract}
The spontaneous breaking of of translational invariance in 
 non-commutative  self-interacting scalar field theory in two dimensions 
is investigated by effective action techniques. The analysis confirms the
existence of the stripe phase, already observed in lattice
simulations, due to the non-local nature of the non-commutative dynamics.
\end{abstract}
 \pacs{11.10.Nx, 11.30.Qc}
 \maketitle

The existence of a phase with conventional long range order, or spontaneous symmetry 
breaking (SSB) for two dimensional (2D) systems with  continuous symmetry group, is 
precluded by the  Coleman-Mermin-Wagner (CMW) theorem, which has been formulated 
specifically for spin models  in \cite{mermin} and for quantum fields in \cite{coleman}.
In these low-dimensional systems the infrared divergences related to the spin waves
or Goldstone modes are so strong that the long range order is destroyed, so that the 
typical order parameter (or scalar field expectation value) vanishes. 
However, in 2D, as for instance in the XY model, 
it is still possible to have  a Kosterlitz-Thouless phase 
transition \cite{kt}  driven by the presence of topological defects and a
quantum field theory which displays an ``almost'' long range order  \cite{witten}.

For quantum fields, the CMW theorem relies on the hypotesis  
of locality. In fact there are  
known exceptions such  as the Liouville theory \cite{roman1,lautrup}.
Another interesting case which  is certainly relevant for this problem 
is the non-commutative formulation of the  quantum field theory
because in this framework the above hypotesis is not respected.

In the non-commutative theory the canonical commutator among 
space-time coordinates is
\be
[x_\mu,x_\nu]= i \theta_{\mu \nu},
\ee
and the product of field operators is non-local, being  defined by the Moyal 
product \cite{connes}. 
For example for the scalar $\lambda \phi^4$ theory the non-commutative  interaction
lagrangian is

\be
L_I = {\lambda 
\over 4!} \phi ^{4*}
\ee
where the Moyal (star) product is defined by ($i,j =1,.,4$)
\cite{connes}
\be\phi ^{4*}(x)= \phi(x) * \phi(x) * \phi(x) * \phi(x) =
\exp\left \lbrack{ i \over 2} \sum_{i<j}\theta_{\mu \nu} 
\partial^{\mu}_{x_i} \partial^{\nu}_{x_j}\right \rbrack
\Bigl ( \phi (x_1) \phi (x_2) \phi (x_3) \phi (x_4)\Bigr ) \left |_{x_{i}=x} \right.
\label{ventiquattro}
\ee

The Moyal product induces an infrared-ultraviolet (IR-UV) connection which deeply 
modifies the structure of the theory with respect to the commutative case.
In fact, it is still unclear whether  a consistent non-commutative field theory exists 
in the continuum limit although recent  lattice simulations suggest 
that a consistent non-commutative $U(1)$ gauge theory can be defined \cite{bietenolzzero},
with possible phenomenological implications \cite{noizero,alt,urru,hin}.

An interesting aspect of non-commutative scalar theories is that, in 4D, 
SSB is possible only in an inhomogeneous phase , 
i.e. where the vacuum expectation value of the field is position-dependent \cite{gubser}. 
This phase is called the stripe phase for the peculiar $x$ dependence of the 
order parameter $<\Phi(x)>_0$.
This unexpected result, conjectured and discussed on the basis of the IR-UV connection  
in \cite{gubser}, has then been obtained by an effective action technique 
\cite{noi,romano} and confirmed by lattice simulations \cite{bietenolz1}.

The stripe phase involves the spontaneous breaking of translational invariance 
which, in 2D, should be forbidden according to the CMW theorem.
Gubser and Sondhi \cite{gubser}, 
on the basis of a Brazovski-like {\bf local} effective lagrangian \cite{braz}
which is quartic in momentum and represents a good description of the 
non-commutative effects near the minimum of the particle self-energy, 
generated by the IR-UV connection, exclude the 2D stripe phase, in agreement 
with the CMW theorem. 
Indeed, they find  that the infrared  behavior of the  2D non-commutative theory 
is even more pathological than that  observed in the  commutative case.
 
On the other hand, it has been reported in \cite{catterall} that, 
in  2D lattice simulations of non-commutative  scalar $\lambda \phi^4$ theory, 
the translational invariance is spontaneously broken and another  
numerical experiment,  \cite{bietenolz2},  
with a more efficient algorithm,  essentially 
confirms the existence of the 2D stripe phase.
Therefore the validity of the  CMW theorem for non-commutative theories is still under 
investigation and in this letter we carry on this analysis, by resorting to the same 
functional technique already used in \cite{noi}, which corresponds  to an Hartree-Fock
computation of the effective action. Within this 
approach which, in the commutative case, confirms the validity of the CMW 
theorem \cite{poli}, and according to the approximations considered in the following, 
it is shown that the stripe phase exists also in 2D, due to non-commutativity.

By following \cite{noi}, let us first verify  that,  
for non-commutative $\lambda \phi^4$ theory in 2D 
there is no spontaneous symmetry breaking with a constant
order parameter. The action is 
\begin{equation}
\label{ventitre}
 I(\phi)=\int d^2 x \left (  {1 \over 2} 
\partial_{\mu} \phi ~ \partial^{\mu} \phi
 - {1 \over 2} m^2 \phi ^2 -{  \lambda 
\over 4!} \phi ^{4*}\right )
\end{equation}
and, by assuming a translational invariant propagator
\be\label{propa}
G(x-y) = \int \frac{d^2p}{(2\pi)^2} \; \frac{e^{-ip(x-y)}}{p^2 - M^2 (p)}\;\;,
\ee
the Cornwall-Jackiw-Tomboulis \cite{cjt} effective action in the Hartree-Fock 
approximation in momentum space is given by \cite{noi}
\bea
 &&\Gamma (\phi , G) = {1 \over 2} \int { d^4 p \over {(2 \pi)^4}} (p^2 - m^2) 
\phi (p) \phi (-p)
\nonumber \\
&& - {\lambda \over {4!}} \left [ {\prod _{a=1}^{4} 
\int { d^4 p_a \over {(2 \pi)^4}} \phi (p_a)}
\right ] \delta ^4 
\left (\sum_a{p_a}\right) \exp\left ({i \over 2} p_1 \wedge p_2\right)
\exp\left ({i \over 2} p_3 \wedge p_4\right) \nonumber \\ 
&& + {i \over 2} \delta^4 (0) \int { d^4 p \over (2 \pi)^4} \ln D(p)G^{-1}(p) + 
 {1 \over 2} \delta^4 (0) \int { d^4 p \over (2 \pi)^4} (p^2 - m^2) G(p) \nonumber \\
&& -  { \lambda \over 6} \int { d^4 p \over (2 \pi)^4} 
\int { d^4 q \over (2 \pi)^4}
 \phi(p) \phi(-p) G(q) \left [1 + {1 \over 2} \exp {(i  q \wedge p)}\right ] \nonumber \\
\label{venticinque}  
&& -   { \lambda \over {12}} \delta^4 (0) 
\int { d^4 p \over (2 \pi)^4} \int { d^4 q \over (2 \pi)^4} G(p)G(q)
 \left [1 + {{1 \over 2}} \exp {(i  q \wedge p)}\right ] 
\eea
where $D(p)$ is the 
free propagator with mass $m$  and $q \wedge p = q_\mu \theta^{\mu \nu} p_\nu$.

In euclidean space the coupled minimization equations, $\delta \Gamma / \delta \phi =0$ 
and $\delta \Gamma / \delta G =0$,
for a constant background $\phi_0$, can be written as
\be\label{emme}
M^2(q) = \mu^2 +  \frac{\lambda}{2}\phi_0^2+
\frac{\lambda}{3} I(\sigma) 
 + \frac{\lambda}{6} \int \frac{d^2p}{(2\pi)^2} \frac{1}{p^2 + M^2(p)} e^{i q\wedge p}
\ee
and
\be
0 = \phi_0 \left (\frac{\lambda}{3} \phi_0^2 - M^2(q)\right ) \delta^2(q)
\ee
where, in eq. (\ref{emme}), the bare mass $m$ has been replaced by $\mu$, 
according to the renormalization :
\be\label{rinormass}
m^2=\mu^2 -  \frac{\lambda}{3} \int \frac{d^2p}{(2\pi)^2} \frac{1}{p^2 + \sigma^2},
\ee
and we have defined
\be
\label{isigma}
I(\sigma) = \int \frac{d^2p}{(2\pi)^2}\left [ 
\frac{1}{p^2 + M^2(p)} -  \frac{1}{p^2 + \sigma^2} \right ].
\ee
The parameter $\sigma$ has the role of  infrared cut-off. 

The non-commutative phase factor connects the infrared and ultraviolet 
regions and therefore one needs a self-consistent 
approach. Let us start by noting that , due to the 
strongly oscillating phase factor, for  $q\to \infty $ the last integral
in the right hand side of eq. (\ref{emme}) takes its contribution from the 
region $p\sim 0$  and therefore we can set 
\be\label{emmeasy}
{\rm lim}_{q \rightarrow \infty} M^2(q) \rightarrow M^2_{asy},
\ee
where the constant $M^2_{asy}$ does not depend on $q$, provided that
\be
\label{intfinito}
\int \frac{d^2p}{(2\pi)^2} 
\frac{1}{p^2 + M^2(p)} 
\ee
is finite (as we shall check self-consistently).

Then, in the infrared region (small $q$) where the mentioned integral  
gets contributions only from large values of the variable $p$, 
we can approximate $M^2(p)$ with its asymptotic value given 
in eq. (\ref{emmeasy}) and we get 
\be
{\rm lim}_{q \rightarrow 0} M^2(q) \simeq  \mu^2 +  \frac{\lambda}{2}\phi_0^2+
\frac{\lambda}{3} I(\sigma) + \frac{\lambda}{6} \int \frac{d^2p}{(2\pi)^2} 
\frac{1}{p^2 + M^2_{asy}} e^{i q\wedge p}
\ee
For  $\theta^{\mu \nu}$ of  maximal rank and eigenvalues $\pm \theta$, 
 it turns out \cite{gubser} that
\be\label{cappazero}
\int \frac{d^2p}{(2\pi)^2} \frac{1}{p^2 + M^2_{asy}} e^{i q\wedge p} =
\frac{1}{2\pi} K_0(M_{asy} |q| \theta)
\ee
where $K_0$ is the modified 
Bessel function which, for $|q| \rightarrow 0$, has the asymptotic behavior
\be\label{loga}
 K_0(M_{asy} |q| \theta) \rightarrow - {\rm ln} (M_{asy} |q| \theta/2)
\ee
Therefore , for any value of $\sigma$, 
\be
\label{asintoto}
{\rm lim}_{q \rightarrow 0} M^2(q) \simeq  - {\rm ln} (M_{asy} |q| \theta/2)
\ee
which is inconsitent with the second minimization equation
\be
M^2(0)= \frac{\lambda}{3}\phi_0^2
\ee
for any finite value of $\phi_0$ \cite{foot}.

This shows that the class of solutions considered so far cannot fulfill 
the minimization equations derived above. The simplest generalization consists
in releasing the constraint of a translationally invariant vacuum expectation value of 
the field which, as seen in \cite{noi} provides a viable solution in the four-dimensional 
problem. In fact we look for solutions in the form of  oscillating field
\be\label{coseno}
\phi(x)= A \;{\rm cos}( Q \cdot x)
\ee
where $A$ is a scalar and  $Q$ a  bidimensional vector,
 which would require a non-translational invariant full propagator $G$. 
However, as discussed in \cite{noi},
a self-consistent approach in  evaluating the effective action is obtained,  
for small $Q$, by a translational invariant ansatz ( see eq. (\ref{propa}) )
where, however, the function $M(p)$ takes into account  the asymptotic,  
infrared and ultraviolet, behaviors
of the gap equation with the non-uniform background. 
The standard homogeneous case, $Q=0$,  is recovered
by neglecting the non-commutative effects that is by 
considering  the so called planar limit, 
$\theta \Lambda^2 \rightarrow \infty$ , where $\Lambda$ is an ultraviolet cut-off. 
From this point of view
a small $Q$ ( in cut-off units) is  associated with  large, but finite, 
values of $\theta \Lambda^2 $ that we shall assume in the remaining part of the paper.

The constant $A$, the  vector $Q$ and the momentum dependent mass 
$M(p)$ are the three parameters that must be determined by the extremization 
of the action in eq. (\ref{venticinque}). Without loss of generality we can choose
a specific direction for the vector $Q$, e.g.: $Q_1=0, \; Q_2=Q$ and then the 
extremum equations for the action read:
\be\label{mini1}
M^2(q) = 
\mu^2+ A^2 {\lambda \over 6}\left (1 + {1 \over 2} \cos {(q_1\theta Q)}\right)+
\frac{\lambda}{3} I(\sigma) 
+ \frac{\lambda}{6} \int \frac{d^2p}{(2\pi)^2} \frac{\cos {[\theta(q_1 p_2-q_2 p_1)]}
}{p^2 + M^2(p)}
\ee

\begin{equation}
\label{mini2}
A\left \lbrack  Q^2 + A^2 {\lambda \over 8} +\mu^2 
+ {\lambda \over 3}I(\sigma)  +
+ \frac{\lambda}{6} \int \frac{d^2p}{(2\pi)^2} \frac{\cos {(p_1 \theta Q)}
}{p^2 + M^2(p)}
\right\rbrack=0
\end{equation}

\begin{equation}\label{mini3}
Q^2  -  {\lambda \over 12} 
\int \frac{d^2p}{(2\pi)^2} \frac{(p_1 \theta Q) \;\sin {(p_1 \theta Q)}
}{p^2 + M^2(p)}=0
\end{equation}

Since the self-consistent approach requires small $Q$
and in particular in the planar theory limit, i.e. for large $\theta$, 
$Q$ must approach zero in such a way that $\theta Q << 1$ ( in cut-off units),
then in eq. (\ref{mini3}) one can approximate $M^2(p) \simeq M^2_{asy}$ to obtain 
\be
\label{qval}
Q^2= \frac{\lambda}{24 \pi} M_{asy}\; \theta \;Q\;\; K_1(M_{asy} \theta Q)
\ee
where $K_1(x)$ is the modified Bessel function.
Also by this semplification, there is no way to solve analitically the coupled 
equations for $A$ and $M(p)$ and therefore
we consider a Raileigh-Ritz approach, i.e. a parametric  ansatz for $M(p)$ and an  
evaluation of the effective action for 
different values of the parameters.

According to the previous analysis, the ansatz for $M(p)$ ,
consistent with the infrared and ultraviolet behaviors of the gap equation, is given by
\be
M^2(p)= M^2_0 - \frac{\lambda}{12 \pi} \ln\left(\frac{|p|}{Q}\right) \phantom{......} |p|< Q
\ee
and
\be
M^2(p)= M^2_0  \phantom{....} |p| \ge Q
\ee
where $M_0$ is a constant. Therefore $Q$ is obtained by eq. (\ref{mini3}) 
with $M_{asy}=M_0$.

We compute the effective action  for the specific field configuration 
given in eq. (\ref{coseno}) and subtract the constant corresponding to the same 
effective action evaluated at $A=0$,  $M^2(p)=\sigma^2$ and $\theta\Lambda^2\to \infty$
(planar limit).
In Figs.1 and 2 we plot the subtracted effective action $W$ as a function of $A$, 
for $\lambda=0.1$, $\theta= 10$ , $\sigma=10^{-5}$ and,
 in Fig. 1,  for $\mu^2=-0.1$ and three different values 
of $M_0$: 0.1, 0.34, 0.6, 
while in Fig. 2  for $\mu^2=0.1$ and  $M_0$: 0.1, 0.22, 0.4
(with all dimensionful quantities expressed in units of the UV cutoff $\Lambda$).
For the parameters $\theta$ and $M_0$ used in the  figures, the corresponding value of 
$Q$, derived from eq. (\ref{qval}), is about $Q\sim 2.6 \; 10^{-2}$, which is consistent 
with the condition  $\theta Q << 1$  in cut-off units, discussed above.

These examples show that for sufficiently negative $\mu^2$ a minimum of $W$ is observed
at $A\neq 0$, whereas, by sufficiently increasing $\mu^2$ to positive values, the minimum 
of $W$ is shifted to $A=0$. 
In each figure we have plotted the effective action for three values of $M_0$,
and the curves labelled with $(b)$ correspond to the optimal value $M_0$ 
which provides the lowest value of $W$. By reducing or increasing  $M_0$
with respect to this optimal value, one finds that the minimum of $W$ is increased.
We have checked that the behavior shown in the figures is stable against changes 
of the infrared regulator, i.e. for any  value of $\sigma$
there is a value of $\mu^2$ below which the minimum of $W$ 
is located at $A\neq 0 $.
Furthermore, the structure displayed in Figs. 1 and 2 does not change when 
varying $\lambda$ and $\theta$ in a wide range of values.

 This can also be partially shown  analytically. Indeed, by handling the 
coupled minimum equations for $A$, $Q$ and $M(p)$: (\ref{mini1}), 
(\ref{mini2}) and (\ref{mini3}),  one can show that
\be
\frac{\lambda}{8} A^2 = Q^2 + M^2(Q)
\ee
and therefore a solution exists if  $M^2(Q)  \ge 0$ , that is if the gap equation 
has a real solution. 

Therefore, although it does not provide a formal proof, the previous analysis 
strongly supports the conclusion that the translational invariance is spontaneously 
broken for the  non-commutative scalar field theory  in 2D, 
i.e. there is a minimum of the effective potential for 
$A \ne 0$,$Q \ne 0$ and $M_0 \ne 0$.

As recalled in the introduction a similar phenomenon occurs in the Liouville
 theory with lagrangian
\be
L = \frac{1}{2}(\partial_\mu \phi)(\partial^\mu \phi) - \frac{m^2}{\beta^2}\exp\left(
\beta \phi\right )
\ee
with $\beta > 0$ and $m^2 >0$. In \cite{roman1} it has been suggested that the 
 spontaneous breakdown of spatial translational invariance
occurs in this  model and that one can build a consistent perturbation theory on a 
static, position-dependent background.
Moreover the existence of these non-translationally 
invariant states has been tested by Montecarlo simulations \cite{lautrup}.

In \cite{gubser} the validity of CMW theorem for the non-commutative scalar 
theory in 2D has been shown for a {\bf complex} scalar field, 
where the O(2) invariance implies zero modes,
as seen by using the Brazovskii-like  {\bf local}
effective lagrangian
\be
L_B = \frac{1}{2} k_1|(\partial^2 +p_c^2)\phi|^2 
+  \frac{1}{2} k_2 \phi^2 + \frac{1}{4} k_4 \phi^4 
\ee
where $k_1,k_4 > 0$, $k_2<0$ and $p_c$ is the momentum where 
the self-energy has a minimum.  

This effective lagrangian is a good approximation near the minimum and  
it is quartic in momentum. Therefore, if there are zero 
modes, the infrared behavior of the fluctuactions is 
worse than the standard 2D case, hence enforcing the validity of  CMW theorem. 
However for a single scalar field the non-translationally invariant 
configuration that gives the zero modes, if any, is not easy to build  and 
in this case there is no evidence that  the CMW theorem is still valid.

In conclusion, our opinion is that the theoretical problem of the 
CMW theorem for non-commutative 2D theory is still open but, 
for a single scalar field and without considering an effective lagrangian, 
the indication given in this paper confirms the lattice results that 
the translational invariance is spontaneously broken due to the 
non-commutative dynamics.

\noindent
\vspace{30pt}

{\bf Acknowledgement} \\
The authors thank Roman Jackiw and So Young Pi for useful suggestions and comments.
The work of  PC has been supported by the ``Bruno Rossi" program. PC is grateful to 
the Boston University for kind hospitality.

\eject

\begin{figure}
\epsfig{figure=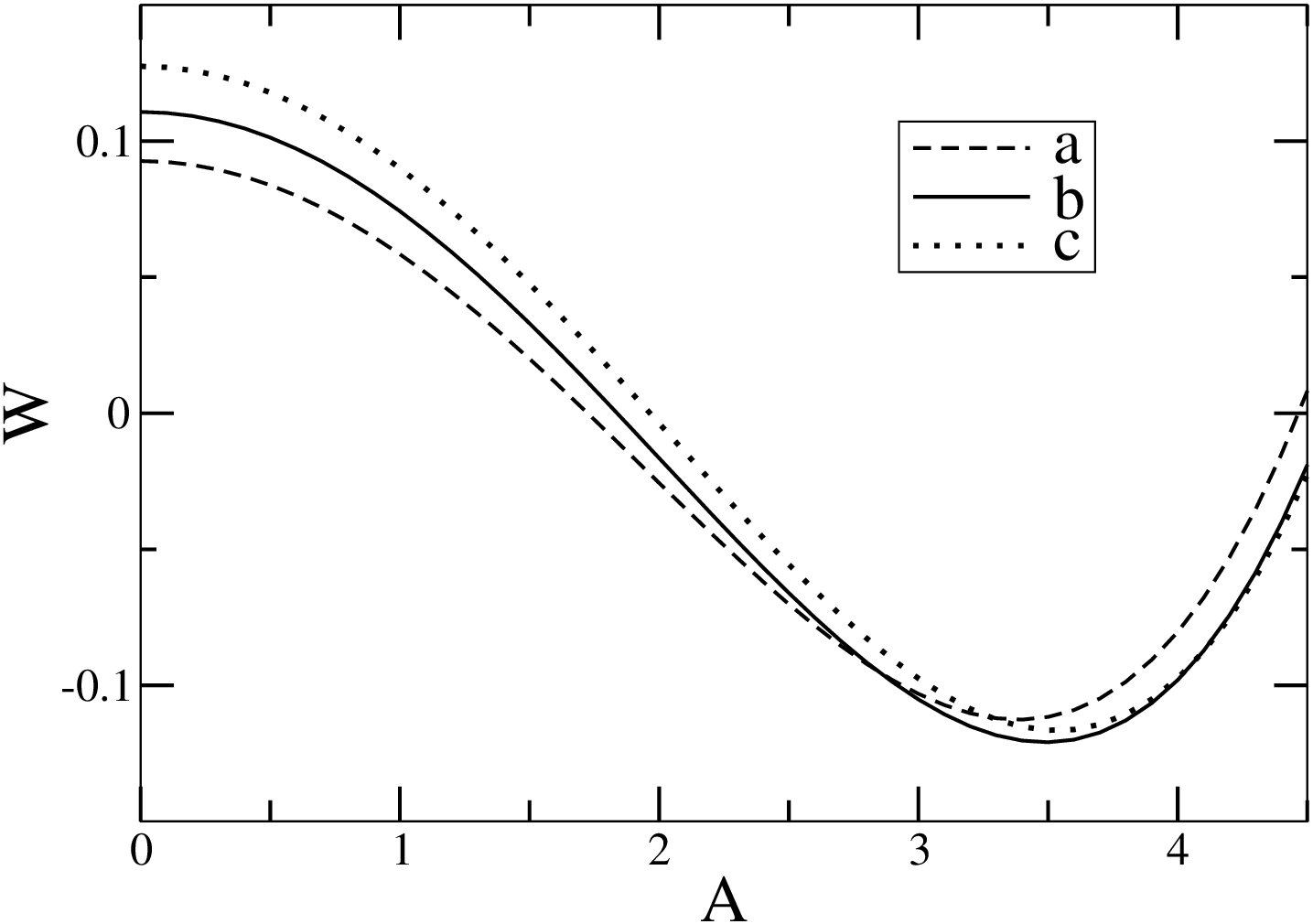,height=12.5 true cm,width=15.5 true cm,angle=0}
\caption{The normalized effective action for $\mu^2=-0.1$ and: 
(a) $M_0=0.1$, (b) $M_0=0.34$, (c)  $M_0=0.6$. }
\end{figure}
\eject
\begin{figure}
\epsfig{figure=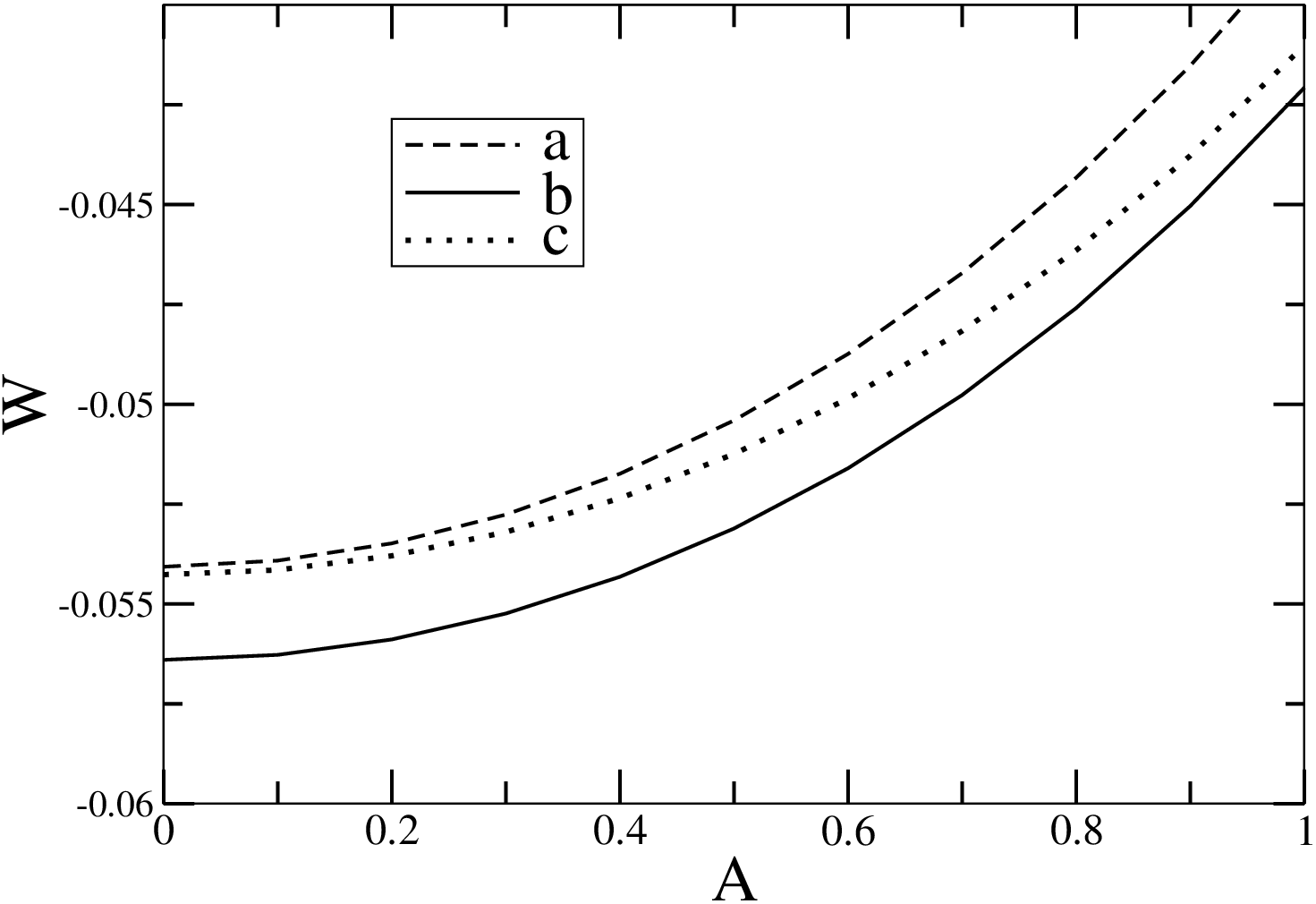,height=12.5 true cm,width=15.5 true cm,angle=0}
\caption{The normalized effective action for $\mu^2=0.1$ and: 
(a) $M_0=0.1$, (b) $M_0=0.22$, (c)  $M_0=0.4$.}
\end{figure}


\begin{thebibliography}{99}
\bibitem{foot} With the ultraviolet and infrared behaviours of eqs. (\ref{emmeasy}) and 
(\ref{asintoto}) the integral in  eq. (\ref{intfinito}) is , selfconsistently, finite.
\end{thebibliography}
\end{document}